# Local optical field variation in the neighborhood of a semiconductor micrograting

Wolfgang Bacsa*, Benjamin Levine, Michel Caumont

Laboratoire Physique des Solides, UMR-CNRS 5477, Université Paul Sabatier, 31062

Toulouse France

Benjamin Dwir

Department of Physics, EPFL, CH-1015-Lausanne, Switzerland

The local optical field of a semiconductor micrograting (GaAs, 10x10μm) is recorded in the middle field region using an optical scanning probe in collection mode at constant height. The recorded image shows the micro-grating with high contrast and a displaced diffraction image. The finite penetration depth of the light leads to a reduced edge resolution in the direction to the illuminating beam direction while the edge contrast in perpendicular direction remains high (~100nm). We use the discrete dipole model to calculate the local optical field to show how the displacement of the diffraction image increases with increasing distance from the surface.

78.68.+m, 78.90.+t

Optical sub-micrometer quantitative imaging of structured surfaces has applications in the field of advanced surface inspection, optical tweezers and photonic crystals. The lateral resolution of lens based systems is limited by the optical wavelength. Near-field optical techniques use a scanning optical probe smaller than the size of the optical wavelength to overcome this limitation. Lens based systems as well as most near-field optical techniques illuminate the sample with a non-planar wave. (1) When averaged in time, the superposition of two counter-propagating waves gives rise to a standing wave or fringe pattern. The superposition of the incident non-planar wave with on the object scattered wave leads in general to a complex fringe pattern. Several theoretical simulations and analytical models have been proposed for a few model systems.(2-4)

Illuminating the sample surface with a collimated monochromatic beam has the advantage that the incident plane wave gives rise to a less complex standing wave pattern.(4-6) When compared to scanning force or scanning tunneling microscopy, optical scanning probe microscopy has the advantage that optical images can be recorded at larger distances from the surface. Although the lateral resolution is reduced at larger distances, regions of interest can be identified before approaching the probe to the surface to record images with a higher lateral resolution at smaller distances. (7)

While an ideal planar and opaque surface illuminated with a plane wave creates a perfectly planar reflected wave front, deviations from this ideal situation are expected to be observable in the case of structured surfaces. The superposition of the incident and reflected waves due to the finite beam width forms a standing wave pattern which is oriented parallel to the surface when averaged in time. Any standing wave pattern can be recorded using a scanning optical probe in collection mode. The deviations from a planar wave are more pronounced in the neighborhood of the surface. This is so because for a given image point the dispersion of the distance to all the contributing points in the surface is larger at smaller distances. This results in a low space coherence which causes the formation of lateral standing waves. At larger distances, the diffraction image increases in size and overlaps with contributions from neighboring regions which increases its complexity. Recording diffraction images in the middle field region has the advantage that a high lateral resolution can be maintained and probe induced effects are reduced while the diffracted image does not overlap with neighboring regions. Here we aim to understand the observed optical contrast of micro-gratings in the middle field region using an optical scanning probe in collection mode.

We have recorded the local field distribution in a plane parallel to the surface at constant height of GaAs micro-gratings (*10x10μm*) created by electron beam lithography (grating grooves depth: *1μm*; grove width 0.7-*1μm*). The optical probe and surface are illuminated by a laser beam with an angle of incidence of *50°* (wavelength: *669nm, 10mW, TE* polarization). We have used bent metal coated optical probes (Nanonics Imaging Ltd., aperture size *100nm*) and a modified scanning probe instrument (Veeco Instruments Inc. AFM Model CP-R). The images were recorded without shear force detection. The incident beam is inclined towards the bottom in all images. Figure 1a) shows the schematic of the used geometry: sample surface (S), the plane of incidence (P), the image plane (I), angle of incidence ($\alpha$) and optical probe. Figure 1.b) shows the recorded image of the micro-grating at a large distance from the sample (*>30μm*). The grating location is indicated by the circle. All experimental images are reproducible and do not depend on a specific optical probe. The image plane is in general not parallel to the surface and cuts through the standing wave, created by the incident and reflected planar waves. The recorded image shows horizontal fringes from the standing waves. The fringe spacing can be used to deduce the tilt angle to correct the substrate orientation. Figure 1.c) shows the same region with the micro-grating after tilt correction which removes the fringes created by the standing wave created by the incident and reflected wave. This clearly shows that the recorded image depends on the orientation of the image plane with respect to the sample. Apart from the micro-grating in the center, other structures (dark regions) and their diffraction fringes are seen on the side. Fine parallel fringes are also seen on the lower side of the location of the micro-grating.

Figure 2.a) shows a recorded optical image (size 20μm) at a smaller distance (<5μm) to the micro-grating. We can distinguish a darker region 1 and a brighter region 2. The grating fringes are seen in both regions. The larger distances from the surface than typically used in nearfield optics leads to a larger phase difference between the incident and scattered field. The interference of the two leads to the formation a diffraction image which is displaced in the direction of the illuminating beam. The fact the size of the micro-grating is larger then the illumination wavelength modifies the reflected wave locally and this reproduces the grating structure in the image plane. The two regions seen in fig. 2.a) are then explained by the superposition of the image formed by the modified reflected wave due to the presence of the grating and the displaced diffraction image of the grating. Figure 2.b) is an enlargement of the lower right corner of region 2 in fig. 2.a). The grating structure with horizontal fringes (image size: *6.25μm*) is clearly seen. We observe that the sharpness of the edges of the grating is different in horizontal and vertical direction. The contrast of the vertical edge groove in horizontal direction is higher. High lateral resolution has been observed earlier (7) using the reflection collection mode. Figure 2.c) shows the same micro-grating rotated, with the grating fringes in vertical orientation, recorded a different image height and keeping the incident beam fixed. The image confirms the high edge resolution perpendicular to the incident beam direction. Displaced diffraction fringes are again superimposed with an image which reproduces the grating structure. Interestingly we see the horizontal fringes of the grating grooves prolong into the vertical edge groove. We believe that the finite penetration of the light into the substrate for GaAs at $\lambda=669nm$ which is *500nm* causes the contrast to spread by 200nm in the direction of the reflected beam and this explains why the contrast is larger in the direction perpendicular to the direction of the incident beam.

In order to have a better understanding of the recorded fringe contrast we have used the discrete dipole model (8-9) to calculate the interference pattern of the scattered field from the micro-grating with the incident field. The model takes into account of the time averaged interference of the incident and scattered electric transverse field component of a single dipole and a plane wave. The longitudinal field component has not been included since the image distance is sufficiently large ($>\lambda$). Furthermore, the scattered field amplitude is several orders of magnitude smaller than the incident field. We therefore neglect the coupling between different discrete dipoles. Higher diffraction orders can be excluded at the image height considered here. The image of the grating is then modelled by the linear superposition of 1180 discrete dipoles.

Figure 3 shows the calculated image contrast at two different image heights ($3\lambda$, $9\lambda$). First we observe that the diffraction image is displaced in the direction of the reflected beam as observed in the experimental image. The two different heights show that the shift of the diffracted image depends on the image height. Second we see that the model calculation reproduces the diffraction fringes around and below the grating. But we note that the contrast is not entirely reproduced. The region 2 is less clearly seen in the simulated image. We attribute the differences between the experimental observations and the numerical simulations to the fact that we use only a two dimensional distribution of dipoles and neglect the three dimensional structure of the grating. The light penetrates into the GaAs substrate which appears to have the effect to enhance the contrast in the recorded image. The finite penetration of the illuminating beam into the substrate is also expected to have the effect of displacing the diffraction image in direction of the reflected beam (10). The calculated image gives the possibility to estimate the upper limit for the distance between surface and image plane. We deduce an upper limit for the image height of 2-3µm from fig. 3.a. The high lateral resolution (80nm) suggests that the image height is smaller and this indicates that the diffraction image is shifted in lateral direction due to the finite penetration of the light into the substrate.

We conclude that the size of the image which reproduces the grating (region 2 Fig 2.b) is limited by the overlap with the diffracted image which also depends on the lateral shift caused by the finite penetration of the light into the substrate. The separation of the grating image and its diffraction image opens the opportunity to image at high lateral resolution at larger distance from the surface where no feedback signal is needed to control the probe in the proximity of the surface. Although the overlap of two images gets smaller, with increasing distance from the surface, the reduced lateral resolution at larger distances, limits the size of objects which can be observed with high lateral resolution on an opaque substrate.

We have recorded constant height images of semiconductor micro-gratings created by electron beam lithography using reflection collection probe optical scanning probe microscopy. Interference fringes due to the tilt of the image plane were corrected by changing the sample orientation. An image of the grating and the superimposed diffracted image are separated in the image plane. The highest observed edge resolution is comparable to the probe aperture. The finite penetration depth of the light leads to a reduced edge resolution in the direction of the illuminating beam. Using the discrete dipole model we have been able to model the diffraction image and explain the displacement of the diffraction image. The larger displacement of the diffraction image observed in the experiment is attributed to the finite penetration of the illuminating beam into the substrate which is not included in our two dimensional model.

\* Current address: Boston University, Electrical and Computer Engineering, 8 Saint Mary's Street, Boston, MA 02215, email: wolfgang.bacsa@lpst.ups-tlse.fr

**Figure 1**

a) Schematic of the reflection collection mode: sample (S), plane of incidence of illuminating beam (P), angle of incidence ($\alpha$), image plane (I) and optical probe. b) Recorded optical image of micro-grating at large distance *(>30μm)*, image size 60x60μm: the incident beam direction is from the lower side and is at an angle of incidence *50°*, fringe spacing of standing wave of incident and reflected beam is *6010nm*. The circle indicates the location of the micro-grating. c) Same scan range and experimental conditions as in b) after changing the tilt of the image plane.

**Figure 2**

Recorded optical image of micro-grating at smaller distance (<5μm) than in figure 1.c): a) image size *20μm*, region 1 is due to diffraction from the micro-grating, region 2 shows the grating fringes. b) Enlargement of the lower right corner (image size *6250nm*) of image a); vertical edges are narrower than horizontal edges. The inset shows a cross section in horizontal direction. c) Rotated micro-grating, same experimental conditions; image size *10μm x 10μm*. The vertical grating edges in the circle are as narrow as in a) and b). The arrow indicates the first diffraction fringe from a dust particle.

**Figure 3**

Calculated image contrast of the micro-grating, using 1180 point dipoles: a) image height *3λ*, the square indicates the location of the grating, region 1 and region 2 are the same as in figure 2; b) image height 9λ, the diffraction fringes are displaced in the direction of the illuminating beam. The displacement depends on the image height. The locations of the dipoles are marked by red points.

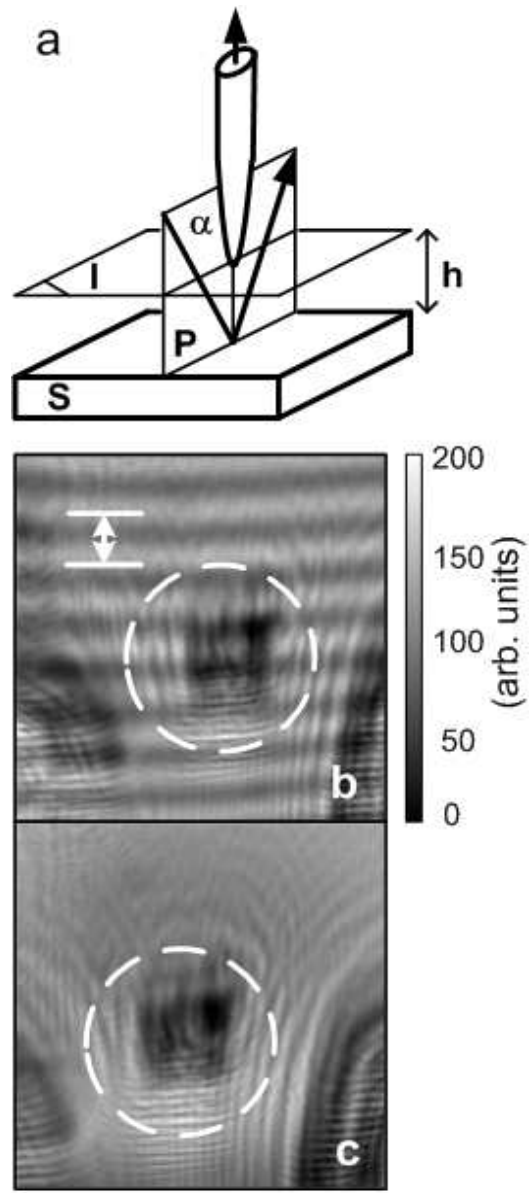

Fig 1

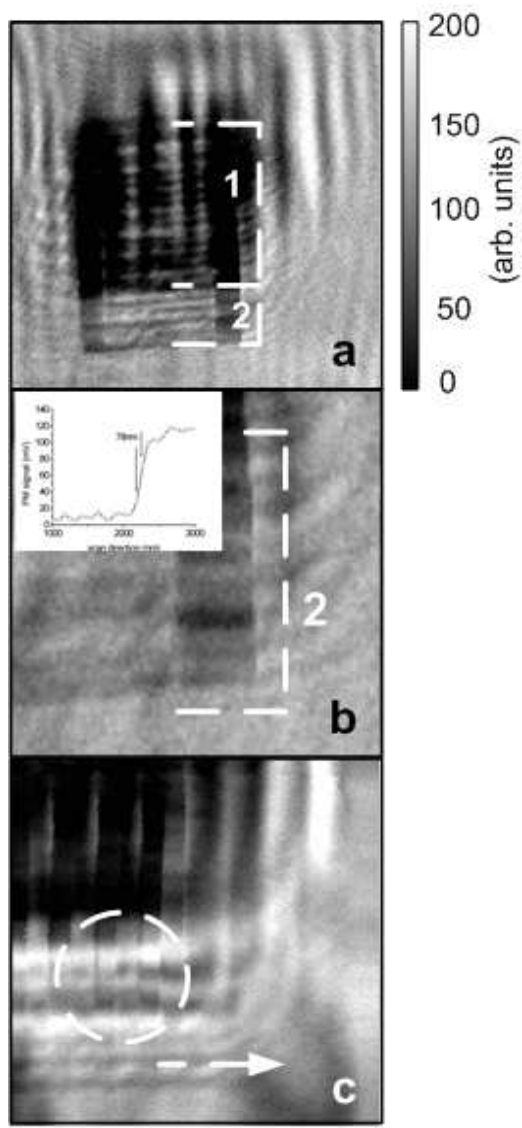

Fig 2

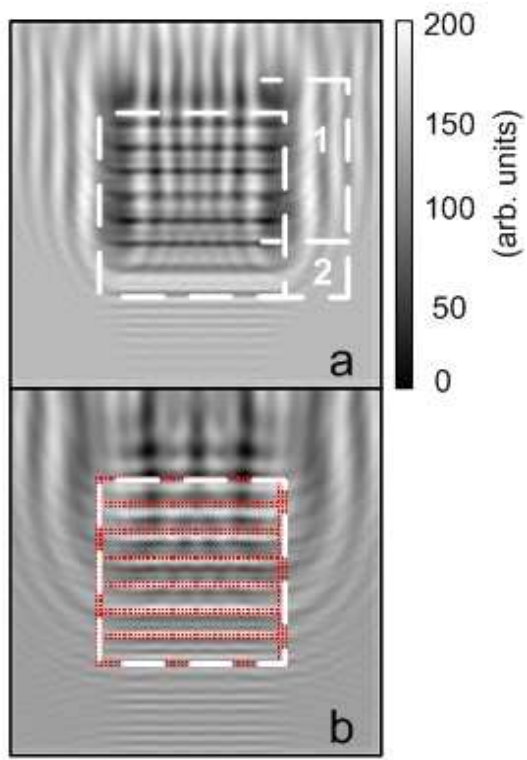

Fig 3